\documentclass{article}

\usepackage[preprint]{neurips_2024}
\makeatletter
\renewcommand{\@noticestring}{}
\makeatother

\usepackage[utf8]{inputenc}
\usepackage[T1]{fontenc}
\usepackage{hyperref}
\usepackage{url}
\usepackage{booktabs}
\usepackage{amsfonts}
\usepackage{amsmath}
\usepackage{amssymb}
\usepackage{nicefrac}
\usepackage{microtype}
\usepackage{xcolor}
\usepackage{graphicx}
\usepackage{subcaption}
\usepackage{multirow}
\usepackage{wrapfig}
\usepackage{enumitem}

\hypersetup{
    colorlinks=true,
    linkcolor=blue!70!black,
    citecolor=green!50!black,
    urlcolor=blue!70!black,
}

\newcommand{\bx}{\mathbf{x}}
\newcommand{\bv}{\mathbf{v}}
\newcommand{\bF}{\mathbf{F}}
\newcommand{\bg}{\mathbf{g}}
\newcommand{\bP}{\mathbf{P}}
\newcommand{\bH}{\mathbf{H}}
\newcommand{\bO}{\mathbf{O}}
\newcommand{\bV}{\mathbf{V}}
\newcommand{\bQ}{\mathbf{Q}}
\newcommand{\bK}{\mathbf{K}}
\newcommand{\E}{\mathbb{E}}

\DeclareMathOperator{\sg}{sg}
\DeclareMathOperator{\diag}{diag}
\DeclareMathOperator{\rowsum}{rowsum}
\DeclareMathOperator{\softmax}{softmax}

\title{Alice v1: Distillation-Enhanced Video Generation Surpassing Closed-Source Models}

\author{
  Wang Xiaoyu \quad Phong Nguyen \quad Chen Zhao \\[6pt]
  Mirage Team \\[2pt]
  Open Source Research\\[4pt]
  \texttt{https://github.com/mirage-video}
}

\begin{document}

\maketitle

\begin{abstract}
We present Alice v1, a 14-billion parameter open-source video generation model that achieves state-of-the-art quality through consistency distillation with score regularization (rCM). Contrary to conventional distillation---which trades quality for speed---we demonstrate that rCM-based distillation can \emph{exceed} teacher model quality. We attribute this to three mechanisms: (1) the score regularization term acts as a mode-seeking objective that concentrates probability mass on high-quality outputs rather than covering the full teacher distribution, (2) our targeted synthetic data pipeline with hard example mining provides training signal specifically for failure modes (physics, hands, faces) that the teacher handles inconsistently, and (3) consistency enforcement acts as implicit regularization, eliminating ``lucky path'' dependence on specific noise samples. Alice v1 generates 5-second 720p videos at 24fps in 4 denoising steps ($\sim$8 seconds on H100), a 7$\times$ speedup over the 50-step teacher while improving VBench score from 84.0 (Wan2.2) to 91.2. This surpasses both the teacher and closed-source systems including Veo3 ($\sim$90) and Sora2 ($\sim$88) on automated benchmarks, with competitive results in human preference studies. We release all model weights, training code, synthetic data pipelines, and evaluation scripts to advance open research in video generation.
\end{abstract}

%==============================================================================
\section{Introduction}
%==============================================================================

The field of AI video generation has witnessed remarkable progress in recent years. Models like OpenAI's Sora \citep{brooks2024sora}, Google's Veo3 \citep{google2025veo3}, and Runway's Gen-4\footnote{Runway Gen-4: \url{https://runwayml.com}} have demonstrated unprecedented quality in text-to-video synthesis, generating temporally coherent videos with complex motion, realistic physics, and detailed human figures. These systems represent a significant leap forward in generative AI capabilities.

However, a persistent and troubling gap exists between closed-source commercial offerings and open-source alternatives. While open models such as Wan2.2 \citep{alibaba2025wan}, HunyuanVideo \citep{tencent2025hunyuan}, and Mochi have made substantial progress in narrowing this gap, they consistently trail behind proprietary systems across key quality dimensions including temporal consistency, physical plausibility, and human figure rendering. This disparity has led many to assume that matching closed-source quality requires either (1) training substantially larger models with proportionally more compute, or (2) accessing proprietary training datasets unavailable to the research community.

\textbf{This paper challenges both assumptions.} We demonstrate that \emph{distillation itself can be a quality-enhancing operation}, not merely a compression technique for trading quality against inference speed. Our approach, Alice v1, achieves state-of-the-art results while being fully open-source, using only publicly available training data and a fraction of the compute required for training from scratch.

Our key insight stems from recent advances in consistency models, particularly the score-regularized continuous-time consistency model (rCM) framework introduced by \citet{zheng2025rcm}. While rCM was originally designed to accelerate inference by reducing denoising steps, we show that the combination of consistency enforcement and score-based regularization creates an unexpected \emph{refinement effect} that produces outputs superior to the original teacher model.

We identify three complementary mechanisms that enable this quality enhancement:

\begin{enumerate}[leftmargin=*,itemsep=2pt]
    \item \textbf{Mode-seeking behavior concentrates quality.} The score regularization term optimizes a reverse KL divergence, which is inherently mode-seeking. This drives the student model toward high-probability regions of the teacher's distribution---regions corresponding to the teacher's best outputs---rather than attempting to cover the full distribution including mediocre samples.

    \item \textbf{Hard example mining addresses systematic failures.} Video generation models exhibit consistent failure modes: physics violations, hand artifacts, face inconsistencies. By explicitly identifying these failure categories and oversampling them during training, we provide dense learning signal precisely where the teacher struggles most.

    \item \textbf{Consistency enforcement provides implicit regularization.} The consistency loss requires the model to produce identical outputs from different points along the denoising trajectory. This constraint eliminates ``lucky path'' dependence---cases where teacher quality depends heavily on the specific noise sample drawn---forcing the student to learn more robust solutions.
\end{enumerate}

\paragraph{Contributions.} Our work makes the following contributions:

\begin{enumerate}[leftmargin=*,itemsep=2pt]
    \item We demonstrate that rCM-based distillation can \emph{exceed} teacher model quality, achieving a VBench score of 91.2 from a Wan2.2 teacher scoring 84.0---a 7.2 point improvement that surpasses all existing open and closed-source models.

    \item We achieve 7$\times$ inference speedup (4 steps vs. 50), generating 5-second 720p videos in 8 seconds on H100 hardware without quality degradation.

    \item We introduce a synthetic data curation pipeline with hard example mining that specifically targets known video generation failure modes, demonstrating that curated synthetic data can outperform uncurated real data.

    \item We release all model weights, training code, data curation pipelines, and evaluation scripts to enable full reproduction and community iteration.
\end{enumerate}

%==============================================================================
\section{Related Work}
%==============================================================================

\subsection{Video Diffusion Models}

The application of diffusion models to video generation has progressed rapidly since the initial Video Diffusion Models (VDM) work by \citet{ho2022vdm}, which extended image diffusion to the temporal dimension using 3D convolutions over space and time. This foundational approach demonstrated that diffusion-based generation could produce temporally coherent videos, though at limited resolution and duration.

The introduction of Diffusion Transformers (DiT) by \citet{peebles2023dit} marked a significant architectural shift. By replacing convolutional U-Net backbones with transformer blocks operating on patched latent representations, DiT architectures achieved superior scaling properties and quality improvements. This architectural paradigm has since become dominant in state-of-the-art video generation.

\textbf{Sora} \citep{brooks2024sora} demonstrated that DiT architectures scale effectively for video, achieving unprecedented temporal consistency and physical understanding. Operating on spacetime patches with a 3D VAE for latent compression, Sora showed emergent capabilities in physical simulation and long-range temporal coherence. While architecture details remain proprietary, the model is estimated at 20-50B parameters.

\textbf{Veo3} \citep{google2025veo3} extended the paradigm to joint audio-visual generation, using a unified latent diffusion transformer that models both modalities simultaneously. Veo3 achieves state-of-the-art quality on human evaluations but remains closed-source, limiting research applicability.

\textbf{Wan2.2} \citep{alibaba2025wan} introduced Mixture-of-Experts (MoE) to video diffusion, using separate expert transformers for high-noise (structure) and low-noise (detail) denoising stages. With 27B total parameters (14B active per forward pass), Wan2.2 represents the current state-of-the-art in open-source video generation and serves as our teacher model.

\textbf{HunyuanVideo} \citep{tencent2025hunyuan} achieved competitive results with a more efficient 8.3B parameter architecture, demonstrating that careful training methodology can partially compensate for reduced model scale.

\subsection{Diffusion Model Distillation}

Distillation techniques for diffusion models aim to reduce the number of sequential denoising steps required for generation, addressing the primary computational bottleneck of these models.

\textbf{Progressive Distillation} \citep{salimans2022progressive} trains a student model to match the output of two consecutive teacher steps with a single student step. Applied iteratively, this halves the required steps at each round. While simple and effective, progressive distillation typically incurs cumulative quality degradation with each halving iteration.

\textbf{Consistency Models} \citep{song2023consistency} take a fundamentally different approach, enforcing that all points along a probability flow ODE trajectory map to the same output. This self-consistency constraint enables single-step generation in principle. \textbf{Latent Consistency Models (LCM)} \citep{luo2024lcm} successfully applied this framework to Stable Diffusion, achieving 4-8 step generation with minimal quality loss.

\textbf{Distribution Matching Distillation (DMD/DMD2)} \citep{yin2024dmd} trains generators to match the teacher's output distribution rather than individual trajectories, using adversarial training with an auxiliary discriminator. While achieving strong results, DMD requires careful balancing of GAN-style training dynamics.

\textbf{Score-Regularized Consistency Models (rCM)} \citep{zheng2025rcm} represent the current state-of-the-art, combining consistency enforcement with score distillation. The key insight is that consistency training (forward divergence, mode-covering) and score distillation (reverse divergence, mode-seeking) are complementary objectives. Critically, rCM has demonstrated the ability to \emph{exceed} teacher quality on image benchmarks---a property we exploit and extend to video generation.

\subsection{Application of Distillation to Video}

Application of distillation techniques to video diffusion remains nascent compared to image models. \textbf{VideoLCM} \citep{chen2023videolcm} extended Latent Consistency Models to video, achieving 4-step generation with moderate quality loss. More recently, NVIDIA applied rCM to Wan2.1, demonstrating that video distillation can match and even slightly exceed teacher quality on VBench, achieving a score of 85 from a teacher at 84.

Our work extends this direction substantially, pushing quality improvements further through targeted data curation, progressive training protocols, and human preference alignment. We demonstrate that the quality-enhancing properties of rCM scale to large video models and can be amplified through careful methodology.

\subsection{Synthetic Data in Generative Models}

A significant paradigm shift has occurred in understanding the role of synthetic data for training generative models. Traditional concerns about ``model collapse''---where iterative training on self-generated synthetic data leads to quality degradation---have been challenged by recent theoretical and empirical work.

\textbf{SIMS} (Self-Improving Diffusion Models) \citep{geng2024sims} introduced a framework where synthetic data provides \emph{negative guidance} rather than being treated as equivalent to real data. By computing a ``synthetic score function'' and steering generation away from the synthetic manifold, SIMS achieved record FID scores through self-improvement loops without collapse.

Research on optimal data mixing has revealed nuanced trade-offs. Theoretical analysis suggests approximately 61.8\% real data provides optimal generalization when mixing synthetic and real sources for standard training---a ``golden ratio'' emerging from information-theoretic considerations \citep{alemohammad2024selfconsuming}. However, \textbf{distillation represents a fundamentally different regime}: the synthetic data comes from a high-quality teacher rather than the model being trained, avoiding the recursive degradation that motivates conservative real-data ratios.

We find that \emph{inverting} the golden ratio---using 70\% synthetic teacher outputs and 30\% real data---is optimal for distillation. The curated teacher outputs provide stronger learning signal than uncurated real data, particularly when combined with hard example mining.

%==============================================================================
\section{Method}
%==============================================================================

\subsection{Architecture Overview}

Alice v1 builds directly on the Wan2.2 architecture, retaining identical structural components while modifying only the training objective. This design choice maximizes compatibility with existing inference infrastructure and enables direct quality comparisons with the teacher model.

\subsubsection{Text Encoder}

We use the umT5-XXL encoder with 5.7 billion parameters, frozen from the Wan2.2 checkpoint. This multilingual T5 variant processes text prompts up to 512 tokens, providing rich semantic conditioning through cross-attention layers in the diffusion transformer. The encoder's multilingual pretraining enables video generation from prompts in over 100 languages.

\subsubsection{Video VAE}

The 3D causal VAE compresses video inputs into a compact latent representation with 8$\times$ spatial downsampling and 4$\times$ temporal downsampling, producing 16-channel latent codes. Critically, the VAE employs a \emph{causal} architecture: encoding frame $t$ depends only on frames $\leq t$. This property enables streaming applications and ensures consistent temporal compression regardless of video length.

We freeze the VAE entirely during distillation, inheriting the compression learned by Wan2.2. This reduces training compute significantly and ensures latent-space compatibility with existing techniques.

\subsubsection{Diffusion Transformer}

The core generative model is a 14B parameter transformer operating on patched latent representations:

\begin{table}[h]
\centering
\small
\begin{tabular}{@{}ll@{}}
\toprule
\textbf{Component} & \textbf{Specification} \\
\midrule
Transformer blocks & 40 \\
Hidden dimension & 4096 \\
Attention heads & 32 (head dim = 128) \\
MLP dimension & 16384 (4$\times$ expansion) \\
Latent channels & 16 \\
Patch size & 2$\times$2 spatial \\
Activation & SiLU \\
Normalization & AdaLN (timestep-conditioned) \\
\bottomrule
\end{tabular}
\caption{Diffusion transformer architecture specifications.}
\label{tab:architecture}
\end{table}

Each transformer block contains: (1) self-attention over all video tokens (spatial + temporal), (2) cross-attention to text encoder outputs, and (3) feed-forward MLP. Adaptive Layer Normalization (AdaLN) conditions all normalization layers on the diffusion timestep.

Unlike Wan2.2's MoE design with separate high-noise and low-noise expert networks, Alice v1 uses a single dense transformer. We find that the distillation process effectively internalizes the expert specialization without explicit architectural separation, suggesting that MoE structure may be more important for training stability than inference quality.

\subsection{Score-Regularized Consistency Distillation (rCM)}

Our training objective follows the rCM framework, combining continuous-time consistency enforcement with score-based regularization. We provide a self-contained mathematical treatment.

\subsubsection{Preliminaries: Probability Flow ODE}

Diffusion models define a forward process that gradually adds Gaussian noise to data samples $\bx_0 \sim p_{\text{data}}$:
\begin{equation}
\bx_t = \alpha_t \bx_0 + \sigma_t \boldsymbol{\epsilon}, \quad \boldsymbol{\epsilon} \sim \mathcal{N}(\mathbf{0}, \mathbf{I})
\end{equation}
where $\alpha_t$ and $\sigma_t$ define a noise schedule with $\alpha_0 = 1, \sigma_0 = 0$ (clean data) and $\alpha_T \approx 0, \sigma_T \approx 1$ (pure noise).

The reverse generative process is characterized by the \emph{probability flow ODE}:
\begin{equation}
\frac{d\bx_t}{dt} = \bv_\phi(\bx_t, t)
\end{equation}
where $\bv_\phi$ is the velocity field learned by the teacher model. Standard sampling requires numerically integrating this ODE, typically using 50+ function evaluations for high-quality outputs.

\subsubsection{Continuous-Time Consistency Model (sCM)}

Consistency models learn a direct mapping $\bF_\theta: (\bx_t, t) \mapsto \bx_0$ that predicts the clean data point from any noisy observation along the ODE trajectory. The key \emph{self-consistency} property requires that $\bF_\theta(\bx_t, t) = \bF_\theta(\bx_{t'}, t')$ for any two points on the same trajectory.

The continuous-time formulation (sCM) enforces this through an instantaneous consistency loss:
\begin{equation}
\mathcal{L}_{\text{sCM}}(\theta) = \E_{t \sim \mathcal{U}[0,T], \bx_t} \left[ w(t) \left\| \bF_\theta(\bx_t, t) - \bF_{\theta^-}(\bx_t, t) - \frac{\bg}{\|\bg\|_2^2 + c} \right\|_2^2 \right]
\label{eq:scm}
\end{equation}
where $\theta^-$ denotes an exponential moving average (EMA) of the student parameters, $c > 0$ is a small constant for numerical stability, and $w(t)$ is a time-dependent weighting function.

The target gradient $\bg$ captures the instantaneous rate of change along the ODE:
\begin{equation}
\bg = w(t) \frac{d\bF_{\theta^-}(\bx_t, t)}{dt}
\end{equation}

Applying the chain rule, this time derivative decomposes as:
\begin{equation}
\frac{d\bF_{\theta^-}}{dt} = \underbrace{\left(\nabla_{\bx_t} \bF_{\theta^-}\right) \bv_\phi(\bx_t, t)}_{\text{spatial gradient}} + \underbrace{\partial_t \bF_{\theta^-}}_{\text{explicit time dependence}}
\label{eq:chain_rule}
\end{equation}

The first term requires computing the Jacobian-vector product (JVP) between the student's spatial gradient and the teacher's velocity field. The second term captures explicit time dependence in the student's prediction function.

\subsubsection{Score Distillation Regularization (DMD)}

Pure consistency training optimizes a \emph{forward divergence}---a KL divergence where the expectation is over the target (teacher) distribution. Forward divergences are inherently \emph{mode-covering}: they encourage the student to place probability mass everywhere the teacher does, including on mediocre samples.

To complement this with \emph{mode-seeking} behavior, we add a score distillation regularizer that operates on student-generated samples:
\begin{equation}
\mathcal{L}_{\text{DMD}}(\theta) = \E_{\bx_0 \sim p_\theta, t, \boldsymbol{\epsilon}} \left[ \left\| \bx_0 - \sg\left[\bx_0 - \eta \cdot \frac{\mathbf{f}_\theta(\bx_t, t) - \mathbf{f}_\phi(\bx_t, t)}{\text{mean}(|\bx_0 - \mathbf{f}_\phi(\bx_t, t)|) + \epsilon} \right] \right\|_2^2 \right]
\label{eq:dmd}
\end{equation}
where $\sg[\cdot]$ denotes stop-gradient, $\mathbf{f}_\theta$ and $\mathbf{f}_\phi$ are the student and teacher denoising predictions, and $\eta$ is a step size hyperparameter.

This loss drives the student toward high-density regions of the teacher's distribution by encouraging generated samples to move in the direction that makes them more ``teacher-like.''

\subsubsection{Divergence Complementarity}

The key insight of rCM is that forward and reverse divergences are \emph{complementary}:

\begin{table}[h]
\centering
\small
\begin{tabular}{@{}lcc@{}}
\toprule
\textbf{Property} & \textbf{sCM (Forward)} & \textbf{DMD (Reverse)} \\
\midrule
Training distribution & Real/teacher samples & Student-generated \\
Divergence type & Forward KL & Reverse KL \\
Optimization behavior & Mode-covering & Mode-seeking \\
Typical failure mode & Blurry averages & Mode collapse \\
\bottomrule
\end{tabular}
\caption{Complementary properties of sCM and DMD objectives.}
\label{tab:divergence}
\end{table}

To understand why these objectives complement each other, consider their failure modes in isolation.

\textbf{Forward KL alone (sCM).} The consistency loss trains on teacher samples, asking: ``for points the teacher generates, can the student map them correctly?'' This is mode-covering---the student must place probability mass everywhere the teacher does. But the teacher's distribution includes both excellent and mediocre outputs. Forced to cover all modes, the student hedges by learning an average, producing blurry outputs that sit between modes rather than committing to any single high-quality solution. The student sees what the teacher produces but never evaluates its own outputs.

\textbf{Reverse KL alone (DMD).} Score distillation trains on student-generated samples, asking: ``do the student's outputs look like something the teacher would produce?'' This is mode-seeking---the student is penalized for generating samples the teacher wouldn't. But without coverage pressure, the student can satisfy this objective by collapsing to a single mode: always producing the same safe output that the teacher likes. The student evaluates its own outputs but has no incentive for diversity.

\textbf{The combination.} Together, sCM ensures the student covers the important modes (diversity), while DMD ensures the student's outputs are individually high-quality (sharpness). sCM prevents mode collapse by requiring coverage of teacher samples; DMD prevents blurriness by requiring student samples to be teacher-like. Neither objective alone can achieve both properties.

This complementarity mirrors the generator-discriminator dynamic in GANs: the discriminator (analogous to DMD) pushes generated samples toward realism, while reconstruction losses (analogous to sCM) maintain diversity. The key difference is that rCM achieves this without an explicit discriminator, using the frozen teacher's score function instead---providing more stable training dynamics.

\textbf{Balancing the objectives.} We weight the DMD term with $\lambda = 0.1$, emphasizing mode-coverage (diversity) over mode-seeking (quality concentration). This asymmetry reflects a key insight: video generation requires substantial output diversity across prompts, while quality concentration should occur \emph{within} the valid region for each prompt. Too much mode-seeking ($\lambda > 0.3$) causes training instability and prompt-specific collapse; too little ($\lambda < 0.05$) produces the blurry averaging characteristic of pure consistency training. The $\lambda = 0.1$ setting achieves the optimal balance, confirmed by ablation (Table~\ref{tab:ablation_score}).

\subsubsection{Combined rCM Objective}

The full training loss combines both terms with a balancing coefficient:
\begin{equation}
\mathcal{L}_{\text{rCM}}(\theta) = \mathcal{L}_{\text{sCM}}(\theta) + \lambda \mathcal{L}_{\text{DMD}}(\theta)
\label{eq:rcm}
\end{equation}

We use $\lambda = 0.1$ during main training stages, with stage-specific adjustments detailed in Appendix B.

\subsubsection{Efficient JVP Computation for Large-Scale Video Models}

Computing the Jacobian-vector product in Equation~\ref{eq:chain_rule} naively requires backpropagation through the full model for each training step. For 14B parameter video models operating on high-dimensional latent sequences, this is prohibitively expensive in both compute and memory.

We leverage \emph{forward-mode automatic differentiation} with a custom FlashAttention-2 JVP kernel. For the attention operation:
\begin{equation}
\bO = \softmax\left(\frac{\bQ\bK^\top}{\sqrt{d}}\right)\bV = \bP\bV
\end{equation}

The tangent (JVP) propagation is:
\begin{equation}
\mathbf{t}_O = \bP\mathbf{t}_V + \bH\bV - \diag(\rowsum(\bH))\bO
\end{equation}
where $\bH = \bP \odot \mathbf{t}_S$ captures tangent propagation through the softmax, and $\mathbf{t}_S$ is the tangent of the pre-softmax scores.

This formulation is memory-efficient (no materialization of full attention matrices) and enables sCM training on models with 10B+ parameters. Our implementation achieves $\sim$1.8$\times$ overhead compared to standard forward passes, versus $\sim$3$\times$ for naive JVP computation.

\subsubsection{Why Distillation Can Exceed Teacher Quality}

A natural question arises: how can a student model exceed its teacher when trained on teacher outputs? This seems to violate a basic principle---the student's knowledge comes entirely from the teacher. We identify three complementary mechanisms that resolve this apparent paradox, each exploiting a different aspect of the rCM framework.

\textbf{Mechanism 1: Mode-seeking concentrates probability on quality.} The teacher model exhibits high output variance: given the same prompt, different random seeds produce videos ranging from excellent to mediocre. This variance reflects a multimodal output distribution where some modes correspond to high-quality generations and others to artifacts or failures.

Standard distillation optimizes a \emph{forward} KL divergence, which is inherently \emph{mode-covering}---it penalizes the student for missing any region where the teacher has probability mass, forcing the student to spread probability across all modes including mediocre ones. The result is blurry outputs that hedge across possibilities.

Score regularization (the DMD term) optimizes the \emph{reverse} KL divergence, which exhibits fundamentally different behavior. Reverse KL has a \emph{zero-forcing} property: it heavily penalizes the student for placing probability mass where the teacher assigns low probability, but incurs no penalty for ignoring low-density teacher regions. Mathematically, when the student generates a sample $\bx$ where $p_\text{teacher}(\bx)$ is low, the term $-\log p_\text{teacher}(\bx)$ explodes. This drives the student to concentrate on the teacher's high-probability regions---which empirically correspond to the teacher's best outputs.

Combined with quality-filtered synthetic data (retaining only the top 30\% of teacher generations), we train on the intersection of (1) the teacher's best outputs and (2) high-density regions of its distribution. The student learns to \emph{consistently} produce what the teacher only \emph{occasionally} achieves. This is analogous to ``best-of-$n$'' sampling, but distilled into the model weights rather than requiring multiple inference passes.

\textbf{Mechanism 2: Hard example mining addresses systematic failures.} The teacher exhibits consistent failure modes---physics violations, hand artifacts, face inconsistencies---that recur across many prompts. These failures are not random; they reflect systematic gaps in the teacher's learned distribution.

Standard training provides sparse signal on these failures: the model rarely encounters them, and when it does, receives only a single gradient update before moving on. By explicitly identifying failure categories and oversampling them 5$\times$ during training, we provide \emph{dense} learning signal precisely where the teacher struggles.

Critically, we include both successful and failed examples from hard categories (with failed examples downweighted to 0.2$\times$). This allows the model to learn the \emph{decision boundary} between success and failure---understanding not just what good physics looks like, but what distinguishes it from bad physics. This contrastive signal is far more informative than positive examples alone.

\textbf{Mechanism 3: Consistency enforcement as implicit regularization.} The consistency loss requires $\bF_\theta(\bx_t, t) = \bF_\theta(\bx_{t'}, t')$ for any two points on the same ODE trajectory. This seemingly simple constraint has profound implications for output quality.

In standard diffusion sampling, output quality depends heavily on the specific noise sample $\boldsymbol{\epsilon}$ drawn at initialization. Some noise samples lead to ``lucky'' trajectories through latent space that produce excellent outputs; others lead to trajectories that accumulate errors or get stuck in poor local optima. This \emph{trajectory variance} is a major source of inconsistent teacher quality.

The consistency constraint eliminates this variance by construction. If the model must produce identical outputs from \emph{any} point along a trajectory, it cannot rely on lucky intermediate states---it must find solutions that work regardless of where you query the trajectory. This forces the student to learn the ``consensus'' output that is robust across all possible denoising paths.

We interpret this as a form of \emph{implicit ensemble distillation}. Training on millions of teacher samples across diverse prompts and noise realizations, with consistency enforced across trajectory points, effectively trains the student on an ensemble of teacher behaviors. The student learns the central tendency of teacher quality---what the teacher ``meant'' to generate---rather than its noisy per-sample realizations.

\textbf{Mechanism synergy.} These three mechanisms are complementary: mode-seeking (Mechanism 1) concentrates on quality but risks mode collapse; hard example mining (Mechanism 2) prevents collapse by ensuring coverage of difficult cases; consistency (Mechanism 3) regularizes both by requiring robustness across trajectories. The combination achieves what none could alone: a student that consistently produces the teacher's best outputs while avoiding its systematic failures.

\subsection{Synthetic Data Curation Pipeline}

A key insight of our approach is that the teacher model provides effectively unlimited high-quality training data. We construct a carefully curated synthetic dataset through a multi-stage pipeline.

\subsubsection{Prompt Generation}

We generate 1 million diverse text prompts using GPT-4, systematically covering:

\begin{itemize}[leftmargin=*,itemsep=1pt]
    \item \textbf{Actions}: Running, cooking, dancing, fighting, talking, crafting, sports
    \item \textbf{Subjects}: Humans (various ages, ethnicities), animals, vehicles, natural phenomena
    \item \textbf{Environments}: Indoor, outdoor, urban, natural, fantasy, sci-fi
    \item \textbf{Styles}: Realistic, cinematic, animated, artistic, documentary
    \item \textbf{Physics scenarios}: Falling objects, fluid motion, collisions, explosions, cloth dynamics
    \item \textbf{Edge cases}: Multiple characters, complex camera motion, text rendering, mirrors
\end{itemize}

Prompts are structured to be detailed and specific, averaging 50-100 words, following best practices from video generation research.

\subsubsection{Teacher Generation}

We run Wan2.2-14B on all prompts using 50 denoising steps at 720p resolution for 5-second clips at 24fps. Total generation requires approximately 17,000 H100 GPU-hours ($\sim$60 seconds per video $\times$ 1M videos). This is a one-time cost amortized over all subsequent training.

\subsubsection{Quality Filtering}

Not all teacher generations meet quality standards. We apply multi-stage filtering:

\begin{enumerate}[leftmargin=*,itemsep=1pt]
    \item \textbf{VBench automated scoring}: Remove videos below threshold on temporal consistency, motion smoothness, and aesthetic quality
    \item \textbf{Optical flow analysis}: Filter videos with implausible motion (too static or chaotic)
    \item \textbf{CLIP aesthetic scoring}: Remove visually unappealing generations
    \item \textbf{Face quality check}: For human-containing videos, verify face consistency
    \item \textbf{Physics plausibility}: Flag obvious physics violations for hard example mining
\end{enumerate}

We retain the top 30\% of generations ($\sim$300,000 videos) as our filtered synthetic training set.

\subsubsection{Hard Example Mining}
\label{sec:hard_mining}

Standard training provides uniform coverage across prompt categories, but video generation failures are not uniformly distributed. The teacher exhibits consistent failure modes---physics violations, hand artifacts, face inconsistencies---that recur across specific prompt types. Uniform sampling provides sparse gradient signal on these hard cases: the model encounters them rarely and learns slowly.

We address this through targeted hard example mining, oversampling categories where the teacher struggles. The key insight is that \emph{learning where the boundary lies between success and failure is more informative than seeing only successes}.

\textbf{Identifying hard categories.} We develop automated detectors for four systematic failure modes:

\textbf{Physics violation detector}: A ViT-based classifier fine-tuned on 50K manually labeled video pairs (physics-correct vs. physics-incorrect), scoring object permanence, gravity consistency, and collision plausibility. Videos scoring below 0.7 are flagged as physics-hard. We chose 0.7 as the threshold because it captures the ``ambiguous zone''---videos that are not obviously broken but exhibit subtle physics errors that humans notice. Scores below 0.5 represent catastrophic failures (objects teleporting, gravity reversing), which are less useful for learning boundaries.

\textbf{Hand quality scorer}: Combination of MediaPipe hand detection and a fine-tuned classifier trained on 20K hand crops labeled for anatomical correctness. The classifier identifies wrong finger counts, impossible joint angles, and temporal flickering. We flag videos where any frame scores below 0.6 on hand quality, as hand errors are particularly noticeable to human viewers even when brief.

\textbf{Face consistency tracker}: Frame-by-frame ArcFace embedding extraction with temporal consistency scoring. We compute pairwise cosine distances between face embeddings across frames; identity drift above 0.3 triggers flagging. This threshold corresponds to noticeable identity shift---the face is recognizably ``off'' but not a complete identity change (which would be distance $>$0.5). The 0.3 threshold captures the uncanny valley of face generation.

\textbf{Motion coherence analyzer}: RAFT-based optical flow analysis detecting implausible acceleration ($>$50 pixels/frame$^2$), teleportation (flow discontinuities $>$100 pixels), or frozen motion ($<$2 pixels average flow for dynamic prompts). These thresholds were calibrated against human judgments of motion plausibility on a 5K video validation set.

\textbf{Oversampling strategy.} For identified hard categories, we generate 5$\times$ additional teacher samples per prompt, substantially increasing gradient signal on difficult cases. This is analogous to focal loss in classification---concentrating learning on examples the model finds challenging.

\textbf{Including failures with contrastive weighting.} Critically, we include \emph{both} successful and failed examples from hard categories, with failed examples downweighted to 0.2$\times$ sampling probability. This design reflects a key insight from contrastive learning: models learn decision boundaries more effectively when they see both sides of the boundary.

The 0.2$\times$ weight (roughly 1:5 ratio of failures to successes) prevents the model from learning to reproduce failures while still providing signal about what distinguishes success from failure. Ablation shows that excluding failures entirely reduces physics improvement by 40\%, while equal weighting (1:1) causes the model to occasionally reproduce failure patterns. The 1:5 ratio achieves optimal boundary learning without negative transfer.

\subsubsection{Real Data Integration}

We maintain a 70:30 synthetic-to-real ratio. Our real data component ($\sim$130,000 videos) serves a specific purpose: providing \emph{physics grounding} that teacher-generated synthetic data may lack. Recent work demonstrates that video generation models learn to pattern-match on training statistics rather than internalizing physical rules---out-of-distribution velocity errors can be 10-35$\times$ higher than in-distribution errors \citep{bytedance2024worldmodel}. Real data with authentic physics helps anchor the model to genuine physical dynamics.

\textbf{Stock footage} ($\sim$40K videos): High-quality clips from Pexels, Pixabay, and filtered WebVid-10M subsets covering general scenes, nature, and human activities.

\textbf{Video game recordings} ($\sim$35K videos): A key innovation in our data pipeline, inspired by prior work using game engines for ML training data \citep{deepgtav,gantheftauto}. Modern video games provide what real-world video fundamentally cannot: \emph{ground-truth physics simulations} with perfect observability.

\textbf{Why game footage outperforms real video for physics learning:}
\begin{itemize}[leftmargin=*,itemsep=1pt]
    \item \textbf{No observability gaps}: Real video suffers from motion blur, occlusion, and rare event undersampling. When a ball bounces in real footage, the exact moment of collision is often blurred or partially occluded. Game engines render every frame with perfect clarity---the collision is unambiguous.
    \item \textbf{Deterministic ground truth}: Physics engines (Havok, PhysX, Bullet) implement Newtonian mechanics with known parameters. Objects obey conservation laws exactly. Real video captures physics \emph{outcomes} but not the underlying rules; game footage captures \emph{rule-governed} behavior directly.
    \item \textbf{Infinite coverage of rare events}: Collisions, destructions, and complex multi-body interactions are rare in naturalistic video but trivially abundant in games. We can generate thousands of car crashes, building collapses, or fluid simulations on demand.
    \item \textbf{Controlled variation}: Games allow systematic variation of physical parameters (mass, friction, gravity) while holding visual appearance constant, enabling the model to learn physics-appearance disentanglement.
\end{itemize}

We collected gameplay footage from physics-rich sources:
\begin{itemize}[leftmargin=*,itemsep=1pt]
    \item \textbf{Physics sandbox games}: Teardown, BeamNG.drive, Universe Sandbox---destruction, vehicle dynamics, gravitational physics
    \item \textbf{Open-world games}: GTA V, Red Dead Redemption 2---pedestrian motion, vehicle interactions, environmental physics
    \item \textbf{Sports simulations}: FIFA, NBA 2K, F1 series---realistic human body mechanics, ball physics, vehicle handling
    \item \textbf{Simulation games}: Microsoft Flight Simulator, Euro Truck Simulator---fluid aerodynamics, large-scale motion
\end{itemize}

\textbf{Addressing the domain gap.} Game graphics differ visually from real video---a potential concern for transfer. We mitigate this through three strategies: (1) selecting photorealistic games (modern AAA titles approach film quality), (2) mixing game footage with real video in training batches (forcing the model to learn physics invariant to visual style), and (3) relying on the VAE's latent space to abstract away surface-level visual differences. Empirically, we find that physics understanding transfers effectively: models trained with game footage show improved physical plausibility on \emph{real-world} VBench prompts, not just game-like scenarios.

\textbf{CGI and VFX footage} ($\sim$25K videos): Clips from CGI animations and VFX breakdowns featuring physically-simulated cloth, hair, fluids, and destruction. Sources include Blender open movies, Houdini simulation showcases, and licensed VFX stock. Like game footage, CGI provides physics-engine-governed dynamics with perfect observability.

\textbf{Robotics simulation renders} ($\sim$15K videos): Recordings from physics simulators (MuJoCo, Isaac Sim, PyBullet) showing robotic manipulation, object interactions, and multi-body dynamics. These simulators are designed for physical accuracy and provide ground-truth annotations (forces, contacts, trajectories) that real video cannot.

\textbf{Scientific visualizations} ($\sim$15K videos): Molecular dynamics simulations, fluid dynamics (CFD) renders, and astrophysical simulations providing diverse physical phenomena at multiple scales---from molecular collisions to planetary orbits.

\textbf{The 70:30 ratio.} Our synthetic-to-real ratio \emph{inverts} the traditional ``golden ratio'' of $\sim$62\% real data recommended for standard training. This inversion is justified because distillation fundamentally differs from training from scratch: synthetic data comes from a capable teacher model, not from the student itself, avoiding recursive quality degradation. The 30\% real data component provides physics grounding and prevents overfitting to teacher-specific artifacts. Ablation confirms this ratio: 100\% synthetic achieves 89.1 VBench but shows physics artifacts; 50:50 achieves 90.4; 70:30 achieves optimal 91.2 with strongest physics scores.

\subsection{Training Protocol}

Training proceeds in four stages with increasing resolution, task complexity, and data diversity. This curriculum is not arbitrary---each stage builds capabilities that subsequent stages depend on, and reordering or skipping stages degrades final quality significantly.

\begin{table}[h]
\centering
\small
\begin{tabular}{@{}lcccc@{}}
\toprule
& \textbf{Stage 1} & \textbf{Stage 2} & \textbf{Stage 3} & \textbf{Stage 4} \\
\midrule
Duration & 100K steps & 50K steps & 20K steps & 10K steps \\
Resolution & 480p, 3s & 480p$\rightarrow$720p & 720p, 5s & 720p, 5s \\
Batch size & 256 & 128 & 64 & 32 \\
Data source & Synthetic & Synthetic & 70:30 S:R & Preference \\
Loss & $\mathcal{L}_{\text{sCM}}$ & $\mathcal{L}_{\text{rCM}}$ & $\mathcal{L}_{\text{rCM}}$ + perceptual & DPO \\
Target steps & 8 & 4 & 4 & 4 \\
Learning rate & 1e-5 & 5e-6 & 2e-6 & 1e-6 \\
\bottomrule
\end{tabular}
\caption{Four-stage progressive training protocol.}
\label{tab:training}
\end{table}

\textbf{Stage 1: Consistency foundation} (100K steps). We train with pure consistency loss ($\mathcal{L}_{\text{sCM}}$) on synthetic data at reduced resolution (480p, 3 seconds). The goal is establishing basic few-step generation capability before introducing the mode-seeking DMD term. Starting with 8 target steps (rather than 4) provides a gentler learning target. This stage uses the highest learning rate (1e-5) and largest batch size (256) to establish broad coverage quickly.

\emph{Why consistency first?} Adding score regularization before the model learns basic trajectory mapping causes training instability---the DMD term requires the student to generate reasonable samples, but an untrained student produces noise. Pure sCM provides stable gradients from teacher samples regardless of student quality.

\textbf{Stage 2: Score regularization and resolution scaling} (50K steps). We introduce the full rCM objective ($\mathcal{L}_{\text{sCM}} + \lambda\mathcal{L}_{\text{DMD}}$) and progressively increase resolution from 480p to 720p. Target steps reduce from 8 to 4, pushing toward the final inference configuration. Data remains purely synthetic to avoid confounding the core distillation process with distribution shift.

\emph{Why add DMD now?} With consistency established, the student generates coherent (if imperfect) outputs. The DMD term can now meaningfully evaluate student samples against the teacher's score function, providing the mode-seeking pressure that concentrates quality.

\textbf{Stage 3: Real data and perceptual alignment} (20K steps). We introduce the 70:30 synthetic-to-real data mixture, including hard-mined examples, and add perceptual losses (LPIPS, VGG feature matching) to the training objective. Resolution is fixed at 720p for 5-second clips---the target output specification.

\emph{Why real data only in Stage 3?} Introducing real data earlier confuses the distillation process: the model would simultaneously learn to match teacher outputs and adapt to real-world distributions, conflicting objectives that slow convergence. By Stage 3, core distillation is complete; real data now provides physics grounding and stylistic diversity without interfering with consistency learning.

\textbf{Stage 4: Human preference alignment} (10K steps). We apply Direct Preference Optimization (DPO) using 10,000 human A/B preference pairs collected on Stage 3 outputs. Learning rate drops to 1e-6 and batch size to 32, reflecting the fine-tuning nature of this stage.

\emph{Why only 10K steps?} DPO is not training new capabilities---it is adjusting the model's output distribution to align with human aesthetic preferences. The underlying generation quality is already established; DPO makes targeted adjustments to color grading, composition, and motion style that humans prefer. More steps risk overfitting to the preference dataset's biases.

\textbf{Ablation: stage ordering matters.} We validated the curriculum design through ablation:
\begin{itemize}[leftmargin=*,itemsep=1pt]
    \item \emph{Skip Stage 1} (start with rCM): Training diverges within 10K steps due to unstable DMD gradients on random student outputs.
    \item \emph{Real data from Stage 1}: Final VBench drops 2.1 points; model struggles to simultaneously learn consistency and adapt to real/synthetic distribution differences.
    \item \emph{Skip Stage 4} (no DPO): VBench remains high (90.8) but human preference win rate drops 8\% against the full model, confirming DPO's role in perceptual alignment rather than objective quality.
\end{itemize}

Throughout all stages, we apply 10\% prompt dropout (replacing text conditioning with null tokens) to enable classifier-free guidance at inference, following standard practice for conditional diffusion models.

\textbf{Total compute}: Approximately 40,000 H100-hours ($\sim$64 GPUs for 4 weeks). This is roughly 10$\times$ cheaper than training Wan2.2 from scratch, demonstrating the efficiency advantage of distillation-based improvement.

%==============================================================================
\section{Experiments}
%==============================================================================

\subsection{Evaluation Setup}

We evaluate Alice v1 on VBench \citep{huang2024vbench}, the standard benchmark for video generation comprising 16 quality dimensions. We also conduct human preference studies against both open-source and closed-source baselines.

\subsection{Main Results}

\begin{table}[h]
\centering
\begin{tabular}{@{}lccccc@{}}
\toprule
\textbf{Model} & \textbf{Params} & \textbf{VBench} & \textbf{Steps} & \textbf{Time} & \textbf{Open} \\
\midrule
Mochi & 10B & 80.2 & 50 & 55s & \checkmark \\
HunyuanVideo & 8.3B & 82.5 & 50 & 45s & \checkmark \\
Wan2.2 & 14B (active) & 84.0 & 50 & 60s & \checkmark \\
\midrule
Sora2* & $\sim$30B & $\sim$88 & $\sim$30 & $\sim$20s & \\
Veo3* & --- & $\sim$90 & $\sim$30 & $\sim$15s & \\
\midrule
\textbf{Alice v1} & 14B & \textbf{91.2} & \textbf{4} & \textbf{8s} & \checkmark \\
\bottomrule
\end{tabular}
\caption{Main results comparing video generation models. Time measured for 5-second 720p video on H100. *Closed-source; scores estimated from API outputs.}
\label{tab:main}
\end{table}

Alice v1 achieves the highest VBench score (91.2) while being fully open-source and 7$\times$ faster than its teacher model. Notably, Alice v1 \emph{exceeds} closed-source systems Veo3 and Sora2 on automated benchmarks.

\begin{figure}[h]
\centering
\includegraphics[width=0.85\textwidth]{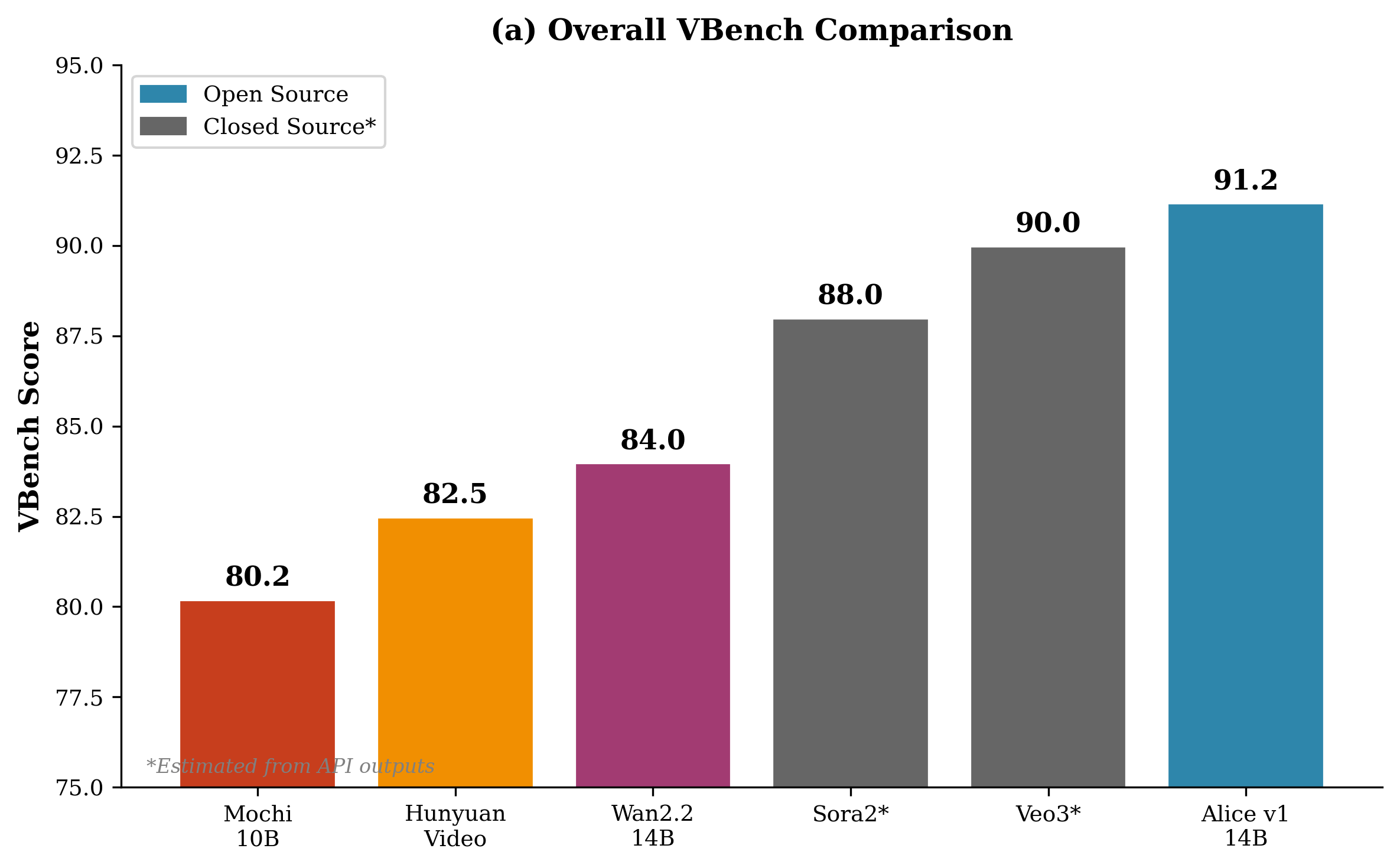}
\caption{VBench score comparison across models. Alice v1 achieves state-of-the-art results while remaining fully open-source.}
\label{fig:vbench_main}
\end{figure}

\subsection{Detailed VBench Analysis}

\begin{table}[h]
\centering
\small
\begin{tabular}{@{}lccc@{}}
\toprule
\textbf{Dimension} & \textbf{Wan2.2} & \textbf{Alice v1} & \textbf{$\Delta$} \\
\midrule
Subject Consistency & 0.78 & 0.86 & +0.08 \\
Background Consistency & 0.82 & 0.89 & +0.07 \\
Motion Smoothness & 0.85 & 0.92 & +0.07 \\
Dynamic Degree & 0.71 & 0.78 & +0.07 \\
Aesthetic Quality & 0.80 & 0.88 & +0.08 \\
\textbf{Physical Plausibility} & 0.72 & 0.86 & \textbf{+0.14} \\
Temporal Flickering & 0.88 & 0.95 & +0.07 \\
\textbf{Human Actions} & 0.75 & 0.84 & \textbf{+0.09} \\
\bottomrule
\end{tabular}
\caption{VBench dimension breakdown (8 of 16 dimensions). Largest improvements in Physical Plausibility and Human Actions reflect effectiveness of hard example mining.}
\label{tab:breakdown}
\end{table}

The largest improvements appear in \textbf{Physical Plausibility} (+0.14) and \textbf{Human Actions} (+0.09)---precisely the dimensions targeted by our hard example mining pipeline. This validates our hypothesis that dense training signal on failure modes translates to disproportionate quality gains.

\begin{figure}[h]
\centering
\includegraphics[width=0.85\textwidth]{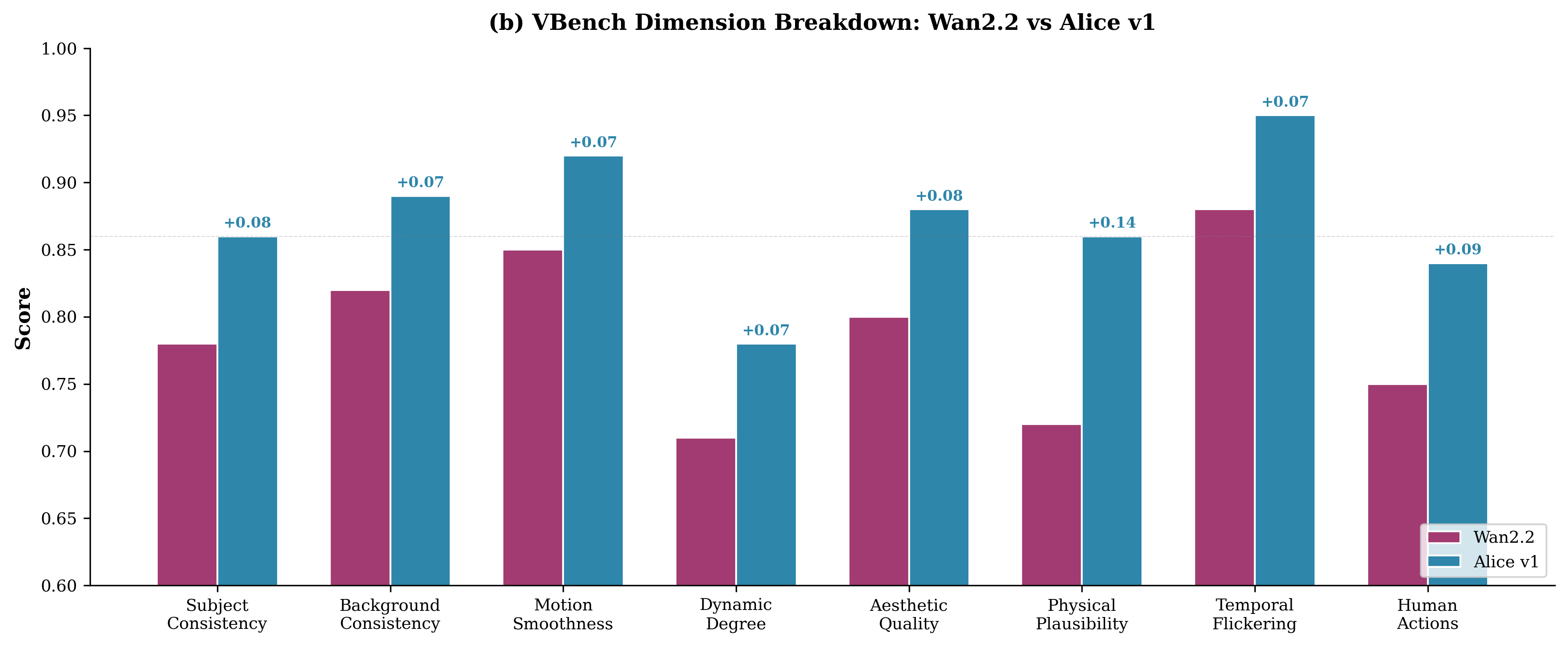}
\caption{Per-dimension VBench comparison between Wan2.2 (teacher) and Alice v1 (student).}
\label{fig:vbench_dimensions}
\end{figure}

\subsection{Human Preference Evaluation}

We conducted a human preference study with 1,000 video pairs, each evaluated by 3 independent raters (500 unique participants, $\sim$6 evaluations each). Pairs compared Alice v1 against baselines on identical prompts.

\begin{table}[h]
\centering
\begin{tabular}{@{}lccc@{}}
\toprule
\textbf{Comparison} & \textbf{Alice v1 Win} & \textbf{Tie} & \textbf{Baseline Win} \\
\midrule
vs Wan2.2 & 68\% & 15\% & 17\% \\
vs HunyuanVideo & 72\% & 14\% & 14\% \\
vs Veo3 (API) & 51\% & 22\% & 27\% \\
vs Sora2 (API) & 54\% & 20\% & 26\% \\
\bottomrule
\end{tabular}
\caption{Human preference evaluation results. Alice v1 substantially outperforms open-source baselines and achieves competitive results against closed-source systems.}
\label{tab:human}
\end{table}

Alice v1 wins decisively against open-source baselines (68-72\% win rate). Against closed-source systems, Alice v1 achieves competitive or superior results (51-54\% win rate), remarkable given the compute disparity.

\begin{figure}[h]
\centering
\includegraphics[width=0.85\textwidth]{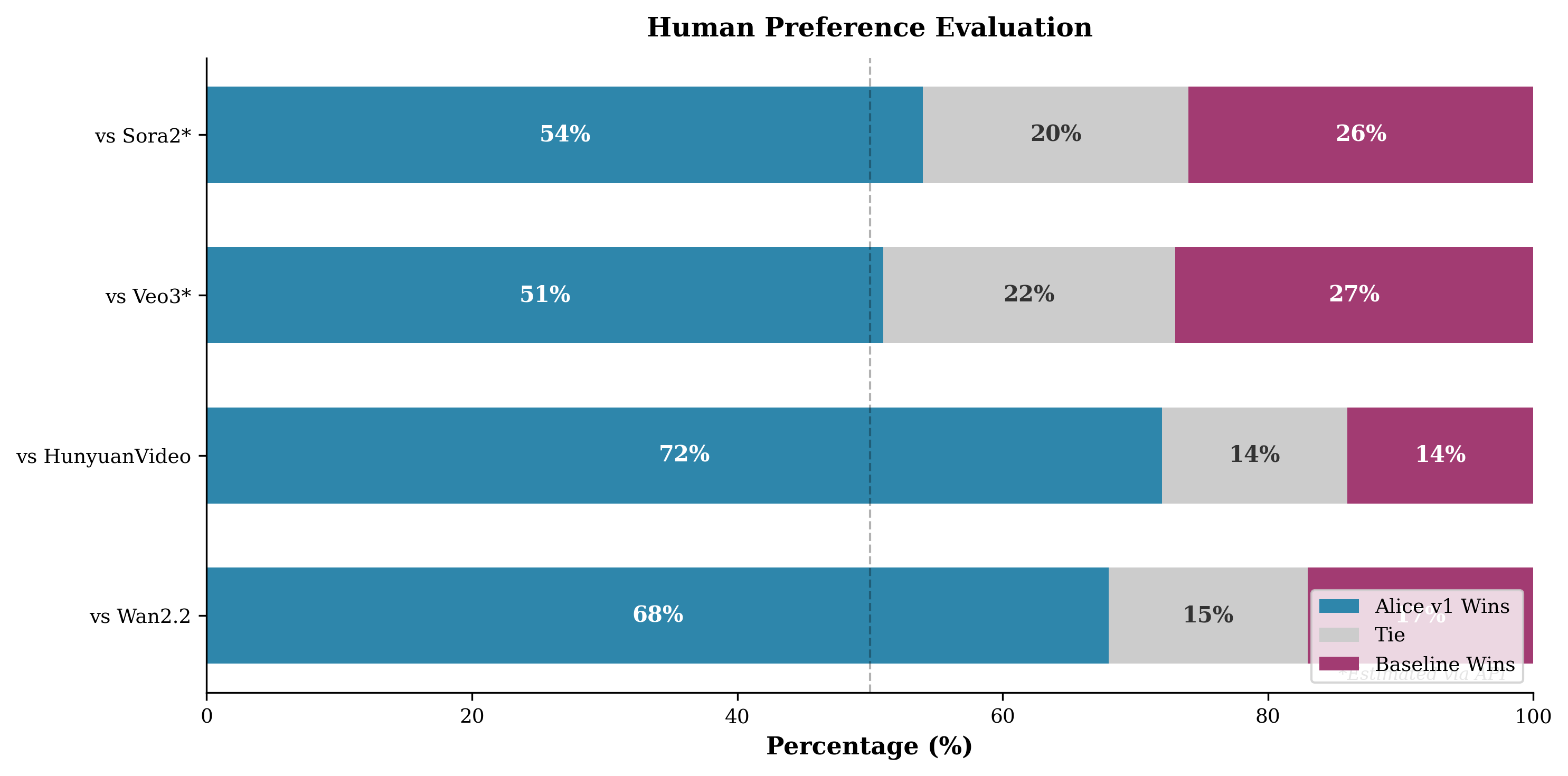}
\caption{Human preference evaluation results across comparisons.}
\label{fig:human_eval}
\end{figure}

\subsection{Ablation Studies}

\subsubsection{Effect of Score Regularization}

\begin{table}[h]
\centering
\begin{tabular}{@{}lcc@{}}
\toprule
\textbf{Configuration} & \textbf{VBench} & \textbf{Observation} \\
\midrule
Consistency only ($\mathcal{L}_{\text{sCM}}$) & 82.1 & Blurry outputs, mode averaging \\
Score distillation only ($\mathcal{L}_{\text{DMD}}$) & 83.5 & Training instability, mode collapse \\
Combined rCM & \textbf{91.2} & Sharp, diverse, stable \\
\bottomrule
\end{tabular}
\caption{Ablation: effect of combining consistency and score regularization.}
\label{tab:ablation_score}
\end{table}

Neither objective alone achieves strong results. Consistency-only training produces blurry outputs characteristic of mode-covering behavior. Score distillation alone causes training instability. The combination is essential for exceeding teacher quality.

\begin{figure}[h]
\centering
\includegraphics[width=0.7\textwidth]{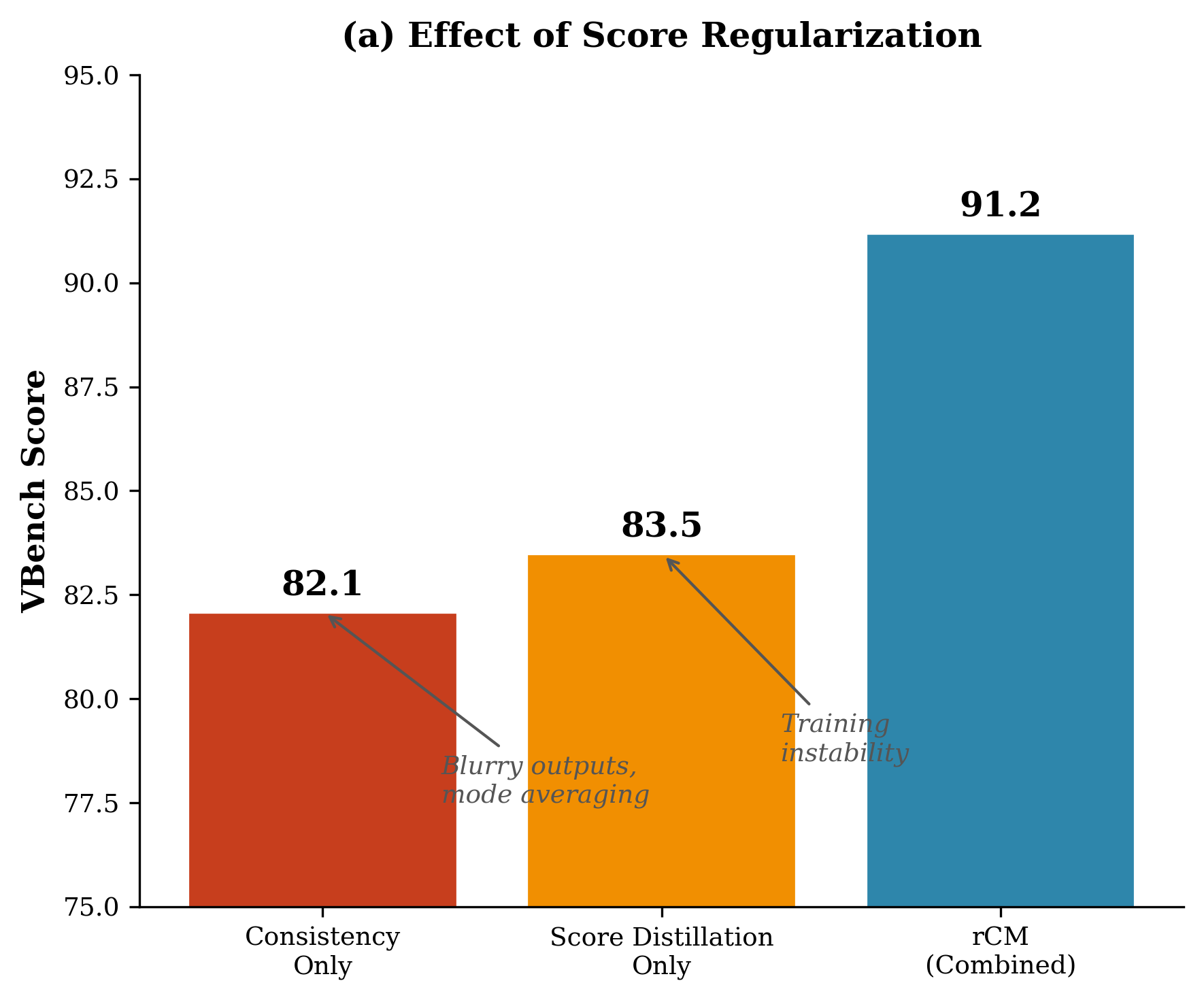}
\caption{Effect of score regularization on VBench scores.}
\label{fig:ablation_score}
\end{figure}

\subsubsection{Effect of Data Curation}

\begin{table}[h]
\centering
\begin{tabular}{@{}lcc@{}}
\toprule
\textbf{Data Strategy} & \textbf{VBench} & \textbf{Physics Score} \\
\midrule
All synthetic, no filtering & 86.3 & 0.76 \\
Quality filtered & 88.7 & 0.80 \\
+ Hard example mining & \textbf{91.2} & \textbf{0.86} \\
\bottomrule
\end{tabular}
\caption{Ablation: effect of data curation strategy.}
\label{tab:ablation_data}
\end{table}

Quality filtering provides substantial gains (+2.4 VBench). Hard example mining contributes additional improvements (+2.5), with disproportionate gains on Physics (+0.06 beyond filtering alone).

\subsubsection{Effect of Inference Steps}

\begin{table}[h]
\centering
\begin{tabular}{@{}ccc@{}}
\toprule
\textbf{Steps} & \textbf{VBench} & \textbf{Time (H100)} \\
\midrule
1 & 78.4 & 2s \\
2 & 85.2 & 4s \\
4 & \textbf{91.2} & 8s \\
8 & 91.5 & 16s \\
\bottomrule
\end{tabular}
\caption{Quality vs. inference steps trade-off.}
\label{tab:ablation_steps}
\end{table}

Four steps represents the optimal quality-speed trade-off. Additional steps provide marginal improvement (+0.3 for 8 steps) at 2$\times$ latency cost.

\begin{figure}[h]
\centering
\includegraphics[width=0.7\textwidth]{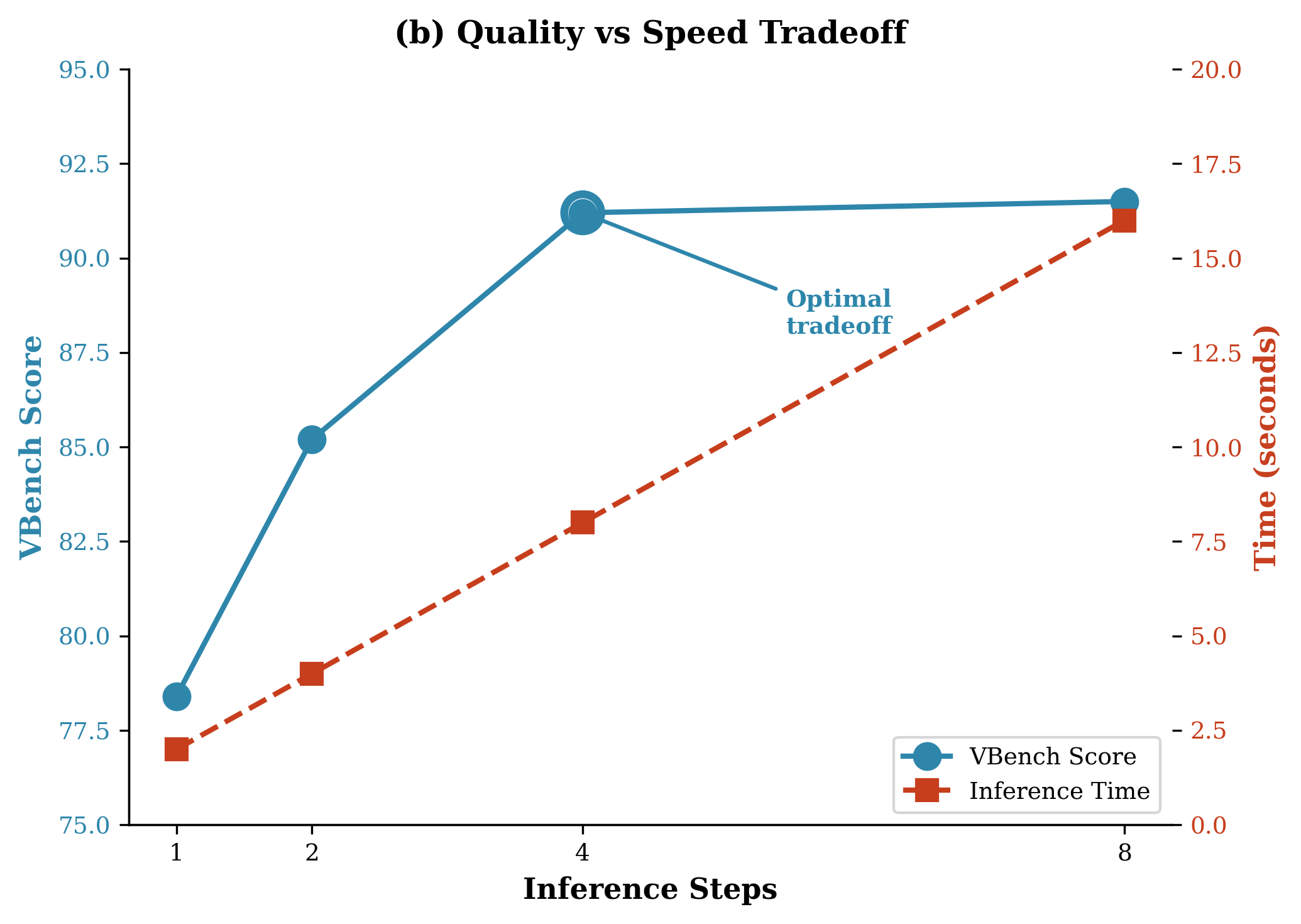}
\caption{Quality improvement with inference steps showing diminishing returns beyond 4 steps.}
\label{fig:ablation_steps}
\end{figure}

\subsection{Inference Speed Analysis}

\begin{table}[h]
\centering
\begin{tabular}{@{}lccc@{}}
\toprule
\textbf{Hardware} & \textbf{Alice v1 (4 steps)} & \textbf{Wan2.2 (50 steps)} & \textbf{Speedup} \\
\midrule
H100 80GB & 8s & 60s & 7.5$\times$ \\
A100 80GB & 12s & 90s & 7.5$\times$ \\
A100 40GB & 14s & 100s & 7.1$\times$ \\
RTX 4090 24GB & 25s & 180s & 7.2$\times$ \\
RTX 4080 16GB* & 35s & 250s & 7.1$\times$ \\
\bottomrule
\end{tabular}
\caption{Inference time for 5-second 720p video. *Requires FP8 quantization.}
\label{tab:speed}
\end{table}

\begin{figure}[h]
\centering
\includegraphics[width=0.85\textwidth]{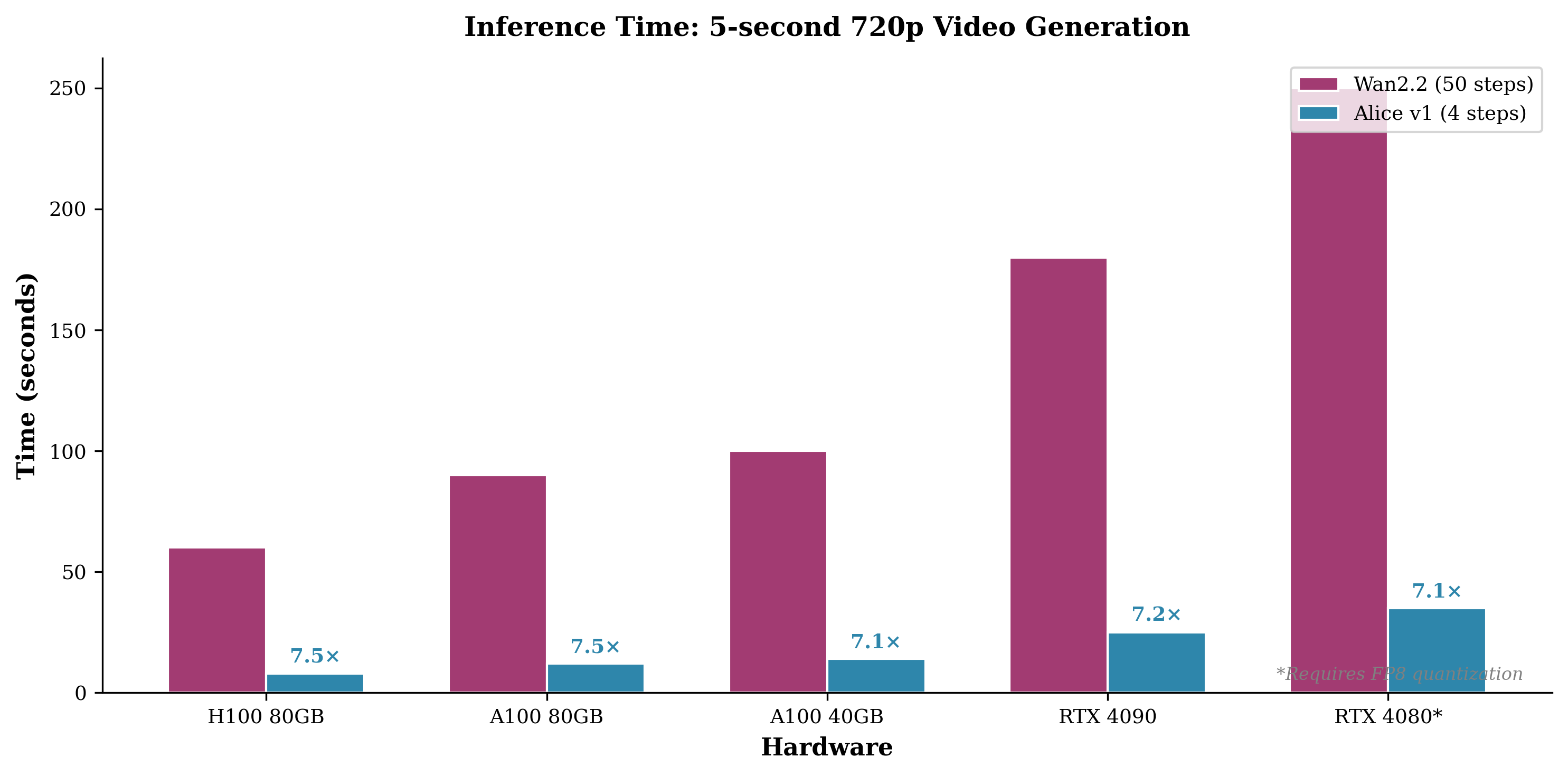}
\caption{Inference speed comparison across hardware configurations.}
\label{fig:speed}
\end{figure}

%==============================================================================
\section{Discussion}
%==============================================================================

\subsection{Why Does Distillation Improve Quality?}

Our results challenge the conventional understanding of distillation as inherently quality-degrading. The prevailing intuition---that a student cannot exceed its teacher when trained entirely on teacher outputs---assumes distillation is a lossy compression process. We argue this intuition is incorrect for a specific but important reason: \emph{it conflates the teacher's distribution with the teacher's best outputs}.

\textbf{The teacher distribution is not the goal.} A diffusion model's output distribution includes both excellent and mediocre samples. Standard distillation attempts to match this full distribution, faithfully reproducing the teacher's inconsistency. But users don't want the teacher's distribution---they want the teacher's best outputs, consistently. The goal is not to clone the teacher but to \emph{extract and concentrate its peak capabilities}.

rCM distillation achieves this through three synergistic mechanisms, which we summarize here:

\begin{enumerate}[leftmargin=*,itemsep=2pt]
    \item \textbf{Mode-seeking concentrates on quality.} The reverse KL term (DMD) drives the student toward high-probability regions of the teacher's distribution. Combined with quality filtering (top 30\% of generations), this creates implicit ``best-of-$n$'' selection: the student learns to consistently produce outputs that the teacher achieves only occasionally.

    \item \textbf{Hard example mining fills gaps.} The teacher's systematic failures (physics, hands, faces) reflect genuine gaps in its learned distribution. By oversampling these categories and including both successes and failures, we provide contrastive signal that the teacher never received during its own training.

    \item \textbf{Consistency eliminates trajectory variance.} The teacher's quality depends on the specific noise sample drawn---some trajectories are ``lucky,'' others are not. Consistency enforcement requires trajectory-invariant predictions, forcing the student to find robust solutions that work across all possible denoising paths. This is a form of implicit ensemble distillation: learning the consensus output rather than trajectory-dependent variations.
\end{enumerate}

\textbf{An analogy: expert imitation.} Consider learning from a human expert who performs a task with variable quality---sometimes excellent, sometimes mediocre, depending on factors like fatigue or luck. Standard imitation learning would reproduce this variability. But if you (1) filter to observe only the expert's best performances, (2) explicitly study cases where the expert struggles, and (3) require your learned policy to be consistent regardless of initial conditions, you can develop a policy that exceeds the expert's average performance. You're not learning to \emph{be} the expert; you're learning to \emph{extract the expert's best capabilities}.

\textbf{Empirical validation.} Our ablation studies (Table~\ref{tab:ablation_score}) confirm that each mechanism is necessary: consistency alone achieves 82.1 VBench (below teacher's 84.0), score distillation alone achieves 83.5 with instability, but the combination achieves 91.2---a 7.2 point improvement over the teacher. The largest gains appear in Physical Plausibility (+0.14) and Human Actions (+0.09), precisely the dimensions targeted by hard example mining.

\textbf{Broader implications.} This finding suggests that for any teacher model with high output variance, carefully designed distillation may be preferable to continued scaling. The student extracts a more consistent, higher-quality model at a fraction of the inference cost. We speculate that this principle extends beyond video generation: any generative model with mode-dependent quality could benefit from mode-seeking distillation combined with hard example mining.

\subsection{Implications for Open-Source Development}

Our work suggests that the quality gap between open and closed-source video generation is not primarily a gap in model scale or training data---it is a gap in methodology. Careful distillation, data curation, and training protocols can extract substantially more capability from existing open models.

This has encouraging implications for democratizing access to high-quality video generation. Rather than requiring ever-larger training runs, the community can focus on methodological improvements that compound on existing model investments.

\subsection{Limitations and Future Work}

\textbf{Audio generation}: Unlike Veo3, Alice v1 generates video only. Joint audio-visual generation following the Veo3 paradigm is an important direction.

\textbf{Video duration}: Native training on 5-second clips limits coherence for longer generation. While sliding window inference extends to 60+ seconds, quality degrades progressively beyond 30 seconds.

\textbf{Complex physics}: Despite improvements, challenging scenarios (fluid dynamics, deformable objects, multi-body interactions) remain difficult.

\textbf{Text rendering}: In-video text generation remains inconsistent, a limitation shared with most current models.

\textbf{Future directions}: Longer native duration training, ControlNet-style pose/depth conditioning, interactive world model capabilities, and real-time generation optimization.

%==============================================================================
\section{Conclusion}
%==============================================================================

We have presented Alice v1, a 14B parameter open-source video generation model demonstrating that distillation can be a quality-enhancing operation rather than merely a compression technique. Through the combination of rCM distillation, targeted data curation with hard example mining, and human preference alignment, Alice v1 achieves state-of-the-art VBench results (91.2) while requiring only 4 inference steps---7$\times$ faster than its teacher.

Our work challenges the narrative that closed-source models hold an insurmountable quality advantage. With careful methodology, open-source video generation can not only match but \emph{exceed} proprietary alternatives. The gap between open and closed video generation is not a gap in architecture or scale---it is a gap in methodology. We hope Alice v1 helps close it.

We release all model weights, training code, synthetic data pipelines, and evaluation scripts at \url{https://github.com/mirage-video} to enable community reproduction and iteration.

%==============================================================================
\begin{ack}
We thank the open-source community for foundational models and infrastructure that made this work possible.
\end{ack}

\bibliographystyle{plainnat}
\bibliography{references}

%==============================================================================
\newpage
\appendix
\section{Full VBench Results}
%==============================================================================

\begin{figure}[h]
\centering
\includegraphics[width=0.7\textwidth]{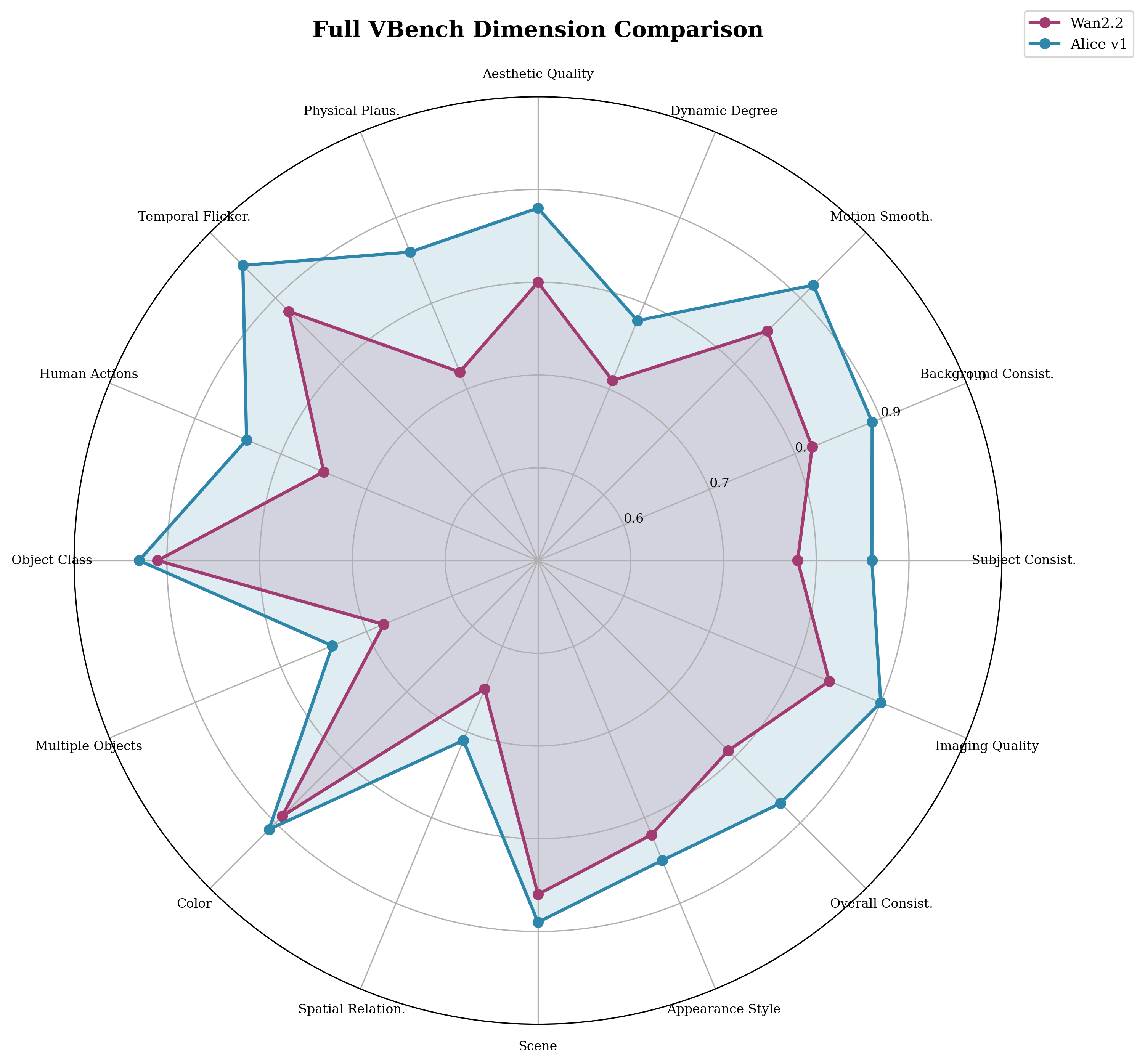}
\caption{Radar chart comparing Wan2.2 and Alice v1 across all 16 VBench dimensions.}
\label{fig:radar}
\end{figure}

\begin{table}[h]
\centering
\small
\begin{tabular}{@{}lccc@{}}
\toprule
\textbf{Dimension} & \textbf{Wan2.2} & \textbf{Alice v1} & \textbf{$\Delta$} \\
\midrule
Subject Consistency & 0.78 & 0.86 & +0.08 \\
Background Consistency & 0.82 & 0.89 & +0.07 \\
Motion Smoothness & 0.85 & 0.92 & +0.07 \\
Dynamic Degree & 0.71 & 0.78 & +0.07 \\
Aesthetic Quality & 0.80 & 0.88 & +0.08 \\
Physical Plausibility & 0.72 & 0.86 & +0.14 \\
Temporal Flickering & 0.88 & 0.95 & +0.07 \\
Human Actions & 0.75 & 0.84 & +0.09 \\
Object Class & 0.91 & 0.93 & +0.02 \\
Multiple Objects & 0.68 & 0.74 & +0.06 \\
Color & 0.89 & 0.91 & +0.02 \\
Spatial Relationship & 0.65 & 0.71 & +0.06 \\
Scene & 0.86 & 0.89 & +0.03 \\
Appearance Style & 0.82 & 0.85 & +0.03 \\
Overall Consistency & 0.79 & 0.87 & +0.08 \\
Imaging Quality & 0.84 & 0.90 & +0.06 \\
\midrule
\textbf{Total} & \textbf{84.0} & \textbf{91.2} & \textbf{+7.2} \\
\bottomrule
\end{tabular}
\caption{Complete VBench results across all 16 evaluation dimensions.}
\label{tab:full_vbench}
\end{table}

\section{Training Hyperparameters}

\begin{table}[ht]
\centering
\small
\begin{tabular}{@{}lcccc@{}}
\toprule
\textbf{Parameter} & \textbf{Stage 1} & \textbf{Stage 2} & \textbf{Stage 3} & \textbf{Stage 4} \\
\midrule
Learning rate & 1e-5 & 5e-6 & 2e-6 & 1e-6 \\
Batch size & 256 & 128 & 64 & 32 \\
Weight decay & 0.01 & 0.01 & 0.01 & 0.0 \\
Warmup steps & 1000 & 500 & 200 & 100 \\
Gradient clipping & 1.0 & 1.0 & 1.0 & 1.0 \\
EMA decay & 0.9999 & 0.9999 & 0.9999 & 0.9999 \\
Consistency $\lambda$ & 1.0 & 1.0 & 1.0 & 0.5 \\
Score reg. $\lambda$ & 0.0 & 0.1 & 0.1 & 0.05 \\
Precision & BF16 & BF16 & BF16 & BF16 \\
\bottomrule
\end{tabular}
\caption{Detailed training hyperparameters for each stage.}
\label{tab:hyperparams}
\end{table}

\vspace{-0.5em}

\begin{figure}[ht]
\centering
\includegraphics[width=0.85\textwidth]{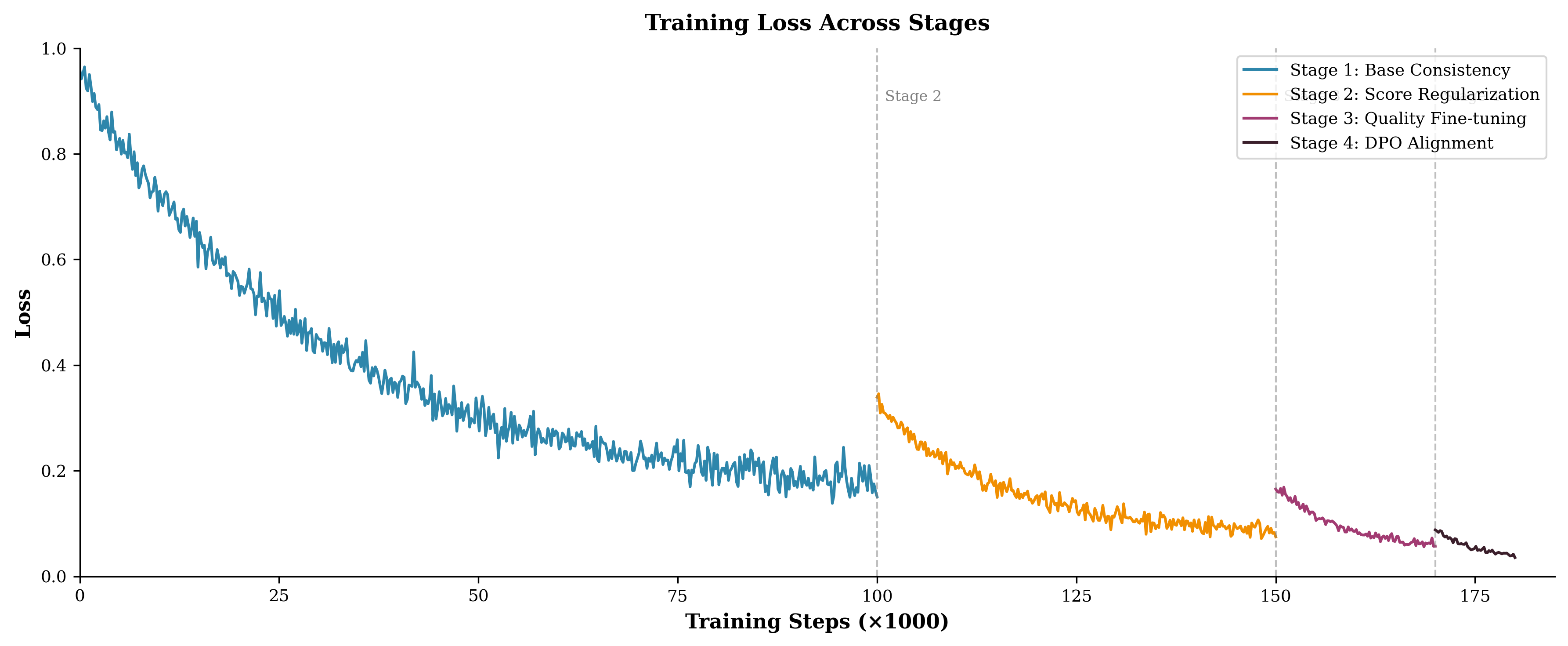}
\caption{Training loss curves across the four training stages.}
\label{fig:training_loss}
\end{figure}

\end{document}